\documentclass[12pt]{iopart}

\newcommand{\rb}{$^{87}$Rb}

\newcommand{\kq}{$^{41}$K}

\newcommand{\kqa}{$^{40}$K}

\newcommand{\ket}[1]{| #1 \rangle}
\newcommand{\prl}{Phys.~Rev.~Lett.}
\newcommand{\pra}{Phys.~Rev.~A}

\newcommand{\hata}{\hat{a}}
\newcommand{\hatb}{\hat{b}}
\newcommand{\hatm}{\hat{m}}
\newcommand{\ebind}{E_{\rm bind}}
\newcommand{\ekin}{E_{\rm kin}}
\newcommand{\ntot}{N_{\rm tot}}
\newcommand{\gammaeff}{\gamma_{\rm eff}}
\newcommand{\be}{\begin{equation}}
\newcommand{\ee}{\end{equation}}
\newcommand{\bea}{\begin{eqnarray}}
\newcommand{\eea}{\end{eqnarray}}

\usepackage{graphicx}  
\begin{document}

\title[]{Collisional and molecular spectroscopy in an ultracold
  Bose-Bose mixture}

\author{G Thalhammer$^1$, G Barontini$^1$, J Catani$^{1,2}$, F
  Rabatti$^1$, C Weber$^1$\footnote{Present address: Institut f\"ur
    Angewandte Physik, Universit\"at Bonn, Wegelerstrasse 8, D-53115
    Bonn, Germany}, A Simoni$^3$, F Minardi$^{1,2}$ and M
  Inguscio$^{1,2}$}

\address{$^1$ LENS European Laboratory for Nonlinear Spectroscopy and
  Universit\`a di Firenze, Via N. Carrara 1, 50019 Sesto Fiorentino,
  Italy} 
\address{$^2$ CNR-INFM, Via G. Sansone 1, 50019 Sesto
  Fiorentino, Italy}  
\address{$^3$ Laboratoire de Physique des Atomes,
  Lasers, Mol\'ecules et Surfaces, UMR 6627 du CNRS and Universit\'e de
  Rennes, 35042 Rennes Cedex, France}
\ead{minardi@lens.unifi.it}

\begin{abstract}
  The route toward a Bose-Einstein condensate of dipolar molecules
  requires the ability to efficiently associate dimers of different
  chemical species and transfer them to the stable rovibrational
  ground state. Here, we report on recent spectroscopic measurements of two
  weakly bound molecular levels and newly observed narrow $d$-wave
  Feshbach resonances. The data are used to improve the collisional model for
  the Bose-Bose mixture \kq\rb, among the most promising candidates
  to create a molecular dipolar BEC. 

\end{abstract}

\maketitle

\section{Introduction}

A new tide in the domain of quantum degenerate gases is rising.
Degenerate dipolar molecules, created from ultracold or degenerate
atoms, are within reach. Such molecules will enable the study of
zero-temperature systems with strong long-range interactions, whereas
degenerate atoms interact substantially only through contact
potentials. Degenerate molecules with dipole-dipole interactions will
provide new quantum phases, will allow the simulation of magnetic spin
systems and will provide candidate qubits for quantum computation.

Ultracold molecules have been produced by several groups, using
either laser cooled atoms in a magneto-optical trap (MOT)
\cite{prl02-pillet} or degenerate atoms
\cite{science02-heinzen}. Due to the energy-momentum conservation,
molecular association cannot be a simple two-body process but requires
a three-body collision, the exchange of photons (photoassociation) or
adiabatic transitions (magnetoassociation).

Molecular Bose-Einstein condensates have already been created
\cite{science03-grimm, nature03-greiner}, but so far only with
homonuclear dimers, whose electric dipole moment is necessarily zero.
On the other hand, heteronuclear molecules have been created starting
from ultracold but not degenerate samples \cite{prl06-sengstock}. A
common challenge to all weakly bound dimer samples is relaxation
decay. While in a MOT production of molecules directly in the
rovibrational ground state has been demonstrated \cite{prl04-gndmot},
the more efficient association from degenerate or quasi degenerate
($T<1\mu$K) atomic samples yields molecules in weakly bound state that
are inevitably unstable to relaxation toward lower lying rovibrational
levels.  Only recently have \kqa\rb\ dimers, which were created by
magnetoassociation, been transferred to the rovibrational ground
state. Their electric dipole moment has been measured to be 0.566
Debye \cite{science08-jin}, in an experiment which represents a
milestone in the route toward degenerate dipolar molecules.

The production of a Bose-Einstein condensate of dipolar molecules,
however, requires association of either two bosons or two
fermions. Therefore we chose to investigate the Bose-Bose mixture
\kq\rb\ in which we associated the first double-species bosonic
molecules \cite{pra08-weber}. Later, bosonic dimers were also created
in a Fermi-Fermi mixture of $^6$Li\kqa\ \cite{prl09-dieckmann}. For
both association and state-transfer, an accurate knowledge of the
interatomic potential is essential. In this work we report
spectroscopic measurements on the \kq\rb\ weakly bound levels,
together with newly observed $d$-wave Feshbach resonances. We use this
new set of data to improve the collisional model for KRb
\cite{pra08-simoni}, earlier adjusted to fit the extensive
observations of Feshbach resonances in the isotopic \kqa\rb\ mixture
\cite{pra06-ferlaino, pra07-arlt}. An accurate knowledge of the
interatomic potential between \kq\ and \rb\ is essential for devising
the transfer of molecules towards the low rovibrational states. The
present work is therefore instrumental to the production of bosonic
KRb molecules with long-range dipolar interactions.

The paper is organized as follows: in \sref{sec:exp} we describe our
experimental procedure, in \sref{sec:exp_results} we present the
data. The theoretical results obtained by the adjustment of the
collisional model are reported in \sref{sec:theo_results}. Finally
we discuss the prospects of a KRb dipolar condensate.

\section{Experimental setup}\label{sec:exp}
Our experimental setup consists of two separate 2-dimensional
magnetooptical traps (MOT), that deliver cold atomic beams of \kq\ and
\rb\ \cite{pra06-catani} into a double-species 3-dimensional MOT.  The
laser cooled mixture is loaded in a quadrupole magnetic trap with a
gradient of 260\,G/cm that is subsequently translated by 31\,mm with a
motorized stage. The quadrupole magnetic field is converted into a
harmonic magnetic trap generated by means of the millimetric trap
described in \cite{pra07-desarlo}. Then we start the microwave
evaporation that expels only \rb\ atoms and cools the mixture
(sympathetic cooling) down to approximately 2\,$\mu$K. At this stage,
the magnetic trap is replaced by a crossed dipole trap generated by
two orthogonal laser beams at 1064\,nm, with waists of $\sim
90\,\mu$m. Once the magnetic potential is completely extinguished,
both species are transferred from the $\ket{F=2,m_F=2}$ to the
$\ket{1,1}$ hyperfine state, by means of consecutive radio-frequency
(rf) ramps in presence of an horizontal bias magnetic field of 7\,G
(adiabatic rapid passage). To guarantee the stability of the mixture
against spin-changing collisions, it is important to transfer the
species with the largest hyperfine splitting first and only later the
second species: in our case we transfer \rb\ first and then \kq. The
adiabatic rapid passages last approximately 30\,ms each and feature
efficiencies comprised from 80 to 95\%.

We raise a homogeneous magnetic field (Feshbach field) along the
vertical direction to access the interspecies Feshbach resonances.  We
calibrate the Feshbach field by measuring the frequency of \rb\
hyperfine transitions. While the field resolution is 20\,mG,
reproducibility is limited to 50\,mG.

The mixture is further cooled by lowering the power of the dipole trap
beams to reach temperatures ranging from 300 to 600\,nK with typically
a few $10^4$ atoms of each species. The mixture is now prepared for
the subsequent experiments.

To associate the molecules and to measure the dimer binding energy, we
modulate the Feshbach field with an additional excitation coil,
driving transitions from unbound to bound pairs. The excitation
frequencies range from 50 to 200\,kHz, but we can drive the
transitions also with half the resonant frequency, effectively
doubling the range of the measurable binding energies. The excitation is
15 to 30\,ms long, with a square amplitude envelope. As explained in
Ref.~\cite{pra08-weber}, we measure our typical excitation amplitudes
to be 130\,mG.

The association of molecules is revealed by the loss of trapped
atoms after the excitation pulse as we scan the excitation
frequency. Such losses, resonant with the excitation frequency, occur
when unbound pairs are converted into weakly bound dimers. Indeed
these dimers are lost from the trap as soon as the subsequent inelastic
collisions rapidly drive them into more deeply bound levels with
a kinetic energy larger than the trap depth.

Likewise, the detection of Feshbach resonances is obtained by the
atomic losses caused by three-body recombination collisions. We
therefore measure the number of atoms remaining after a
fixed time in the dipole trap (hold time), as we scan the Feshbach field
\cite{prl08-gregor}. This method is especially useful for narrow
resonances (such as those of higher partial waves), while it is poorly
accurate in case of broad Feshbach resonances, since the complex
dynamics of atoms and weakly bound dimers must be taken into account
to determine the exact position of Feshbach resonance from the
inelastic losses. When possible, the positions of Feshbach resonances
are better determined from extrapolation to zero of the measured
binding energies.

\section{Experimental results}\label{sec:exp_results}
We report here two different results which are both important for
improving the accuracy of the collisional model for \kq\rb,
namely the measurement of the binding energies of the $s$-wave
Feshbach dimers and the detection of two narrow $d$-wave Feshbach resonances.

\subsection{s-wave molecular levels}
In the vicinity of Feshbach resonances, the binding energy of the
shallowest molecular levels can be accurately measured using rf
spectroscopy. Unbound atom pairs are associated into weakly bound
dimers by means of an oscillating magnetic field ${\mathbf
  b}\cos(\omega t)$.  In our experiment, the ${\mathbf b}$ field is
parallel to the Feshbach field, therefore it represents a modulation
of the latter \cite{prl05-wieman}. The association occurs resonantly
when the oscillation frequency matches the energy difference between
the unbound pair, $E=\ekin$, and the molecular level, $E=-E_b$.

The associated dimers undergo inelastic collisions with the unbound
atoms and decay to more deeply bound levels (``relaxation collisions''). In
this process the binding energy is converted into kinetic energy
that is shared by the collisional partners, which are both expelled from the
trap. Thus, association of molecules can be detected as a resonant
loss of the atomic sample, even if the molecules themselves are too short
lived to be directly detected.

\begin{figure}[tbp]
  \begin{center}
  \includegraphics[width=.6\columnwidth,clip]{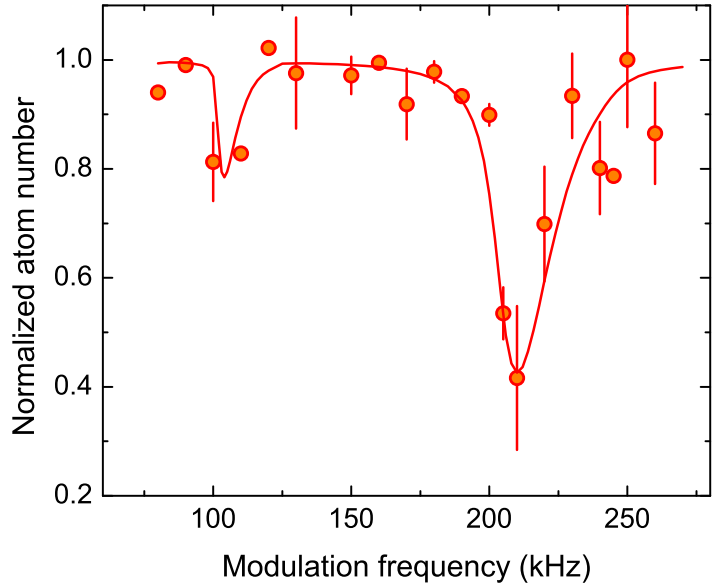}
  \caption{Line shape of rf association of Feshbach dimers around the
    Feshbach resonance at 79\,G: we show the data of the normalized
    total atom number $N_K+N_{\rm Rb}$ remaining after the 15\,ms long
    rf pulse (solid circles), together with the result of numerical integration
    of the model described in Appendix (line). We also observe molecular
    association at half the resonance frequency.}
  \end{center}
\label{fig:rfline}
\end{figure}

In \cite{pra08-weber} we reported the first creation of heteronuclear
bosonic molecules by means of rf association.  We measured the binding
energies of the Feshbach molecular levels next to the Feshbach
resonances at 38 and 79\,G. The atomic losses display a line shape,
shown in \fref{fig:rfline}, that must be modelled to accurately
extract the binding energy. For this purpose, we employ a simple
model, which is described in the Appendix, that captures the salient
features of the line shape. Our analysis reveals that the line shape
is inhomogeneously broadened due to the thermal distribution of
kinetic energies of the unbound atoms. The uncertainty of the measured
resonant frequencies is 1\%, which is mainly due to the fit precision.

In the following, we use these data to improve on the collisional
model which predicts the weakly bound molecular levels near the
scattering threshold and the scattering length as a function of the
magnetic field.

\subsection{Higher order Feshbach resonances} 
\begin{figure}[tbp]
  \includegraphics[width=\columnwidth,clip]{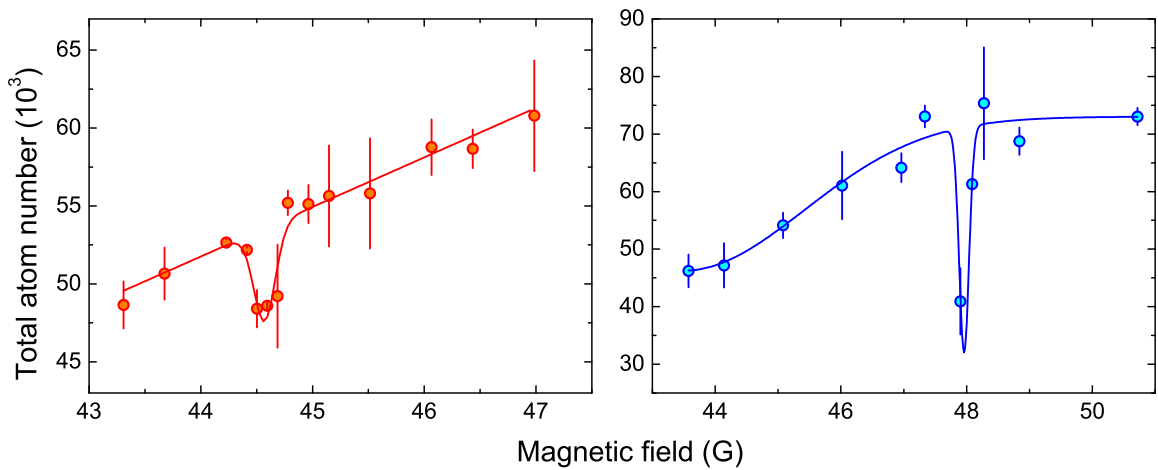}
  \caption{Loss features corresponding to the narrow $d$-wave Feshbach
    resonances. Both features are fitted by a narrow Gaussian
    line shape on a broader pedestal, that is linear in (a) and
    Gaussian in (b). Notice that the data set (b) does not show the
    feature at 44.6\,G since the magnetic field steps are too coarse
    around this value.}
\label{fig:ddwave}
\end{figure}

In addition to the strong losses caused by two $s$-wave Feshbach
resonances, at 38 and 79\,G respectively, we observe two
narrow features, that we attribute to higher partial waves Feshbach
resonances based on the collisional model developed in
\cite{pra08-simoni}. These peaks, shown in \fref{fig:ddwave}, were
observed by simply holding the atomic mixture in the optical trap for 100\,ms
at different magnetic fields.

A first loss feature, detected at a temperature of 400\,nK,  is
centred at a magnetic field of 44.58(0.05)\,G, with a very narrow
half-width at half-maximum (HWHM) of 0.1\,G.

The second loss feature is detected at a temperature of 450\,nK,
centred at a magnetic field of 47.96(0.02)\,G. The width of this peak,
HWHM=0.08\,G, is at the limit of our experimental resolution.  For
both loss peaks the uncertainty on the position is mainly systematic
and equals 0.05\,G.  The assignment of these narrow loss features to
$d$-wave Feshbach resonances follows from the collisional model, as
described in the next section.

\section{Theoretical results}\label{sec:theo_results}

Collisional properties of KRb isotopes are well
understood. In Ref.~\cite{pra08-simoni}, data on $^{39}$K$^{87}$Rb and
\kqa\rb\ Feshbach resonances have been used to construct an accurate
quantum scattering model. Here we add to the data set binding energies
as a function of magnetic field of two near dissociation levels and
magnetic resonance positions in the \kq\rb\ isotopic pair (see
\tref{tab1} and \tref{tab2}). Molecular state energies are particularly
valuable as they are immune to possible systematic shifts affecting
the determination of the two-body resonance locations obtained from
the observed maxima of the three-body recombination rate.

\begin{table}[tbp] \begin{center} \caption{ Comparison of experimental
      $B_{\rm exp}$ and theoretical $B_{\rm th}$ Feshbach resonances
      positions for collisions of \kq\ and \rb\ in their respective
      lowest Zeeman sublevel $|F=1,m_F=1\rangle$. Errors on
      theoretical data have been recalculated from the model.  The
      molecular quantum numbers $(f m_f \ell^\prime)$ associated to
      the observed resonances are reported in the rightmost column.}
    \vskip 12pt
 \label{tab1}
\begin{tabular}{  r@{.}l  r@{.}l   l   c c}
  \hline \hline
  \multicolumn{2}{l}{$B_{\rm exp}$(G)} & \multicolumn{2}{l}{$B_{\rm th}$(G)} 
  & assignment  \\
  \hline
  44&58(5) &  44&63(2)  & $(212)$ \\   %
  47&96(10)&  47&90(6)  & $(202)$    \\   %
  78&57(5) &  78&67(4)  & $(320)$    \\  %
\hline \hline
\end{tabular}
\end{center}
\end{table}

\begin{table}[tbp] 
\begin{center} 
  \caption{Comparison of observed $E_{\rm m,exp}$ and numerically
    calculated $E_{\rm m,th}$ $s$-wave molecular levels as a function
    of the magnetic field $B$.  The rightmost column shows the
    molecular quantum numbers ($f m_f \ell^\prime$).  Errors in the
    $E_{\rm m,exp}$ column have been used for the weighted $\chi^2$
    and account for direct energy error plus the uncertainty 0.05\,G
    in the magnetic field (see text). Errors on theoretical data have
    been recalculated from the model.}  \vskip 12pt
 \label{tab2}
\begin{tabular}{ r@{.}l  r@{}l  r@{}l c  }
\hline \hline
  \multicolumn{2}{l}{$B$(G)} &
  \multicolumn{2}{l}{$E_{\rm m,exp}$(kHz)} & \multicolumn{2}{l}{$E_{\rm m,th}$(kHz)} 
  &  $(f m_f \ell^\prime)$\\
\hline
14&62 & -142&(2) &  -141&.0(9)& (220) \\   %
16&20 & -128&(2) &  -130&.7(8)&    \\   %
16&25 & -132&(2) &  -130&.3(8)&     \\   %
16&80 & -130&(2) &  -126&.6(8)&     \\   %
17&82 & -118&(2) &  -119&.6(8)&     \\   %
19&50 & -107&(2) &  -107&.7(7)&     \\  \hline %
77&83 & -454&(25) &  -400&(16)& (320) \\   %
77&92 & -344&(25) &  -321&(14)&    \\   %
78&01 & -264&(15) &  -250&(13)&    \\   %
78&10 & -194&(15) &  -187&(12)&    \\   %
78&19 & -134&(15) &  -132&(11)&    \\   %
78&24 & -114&(15) &  -107&(10)&    \\   %
78&25 & -114&(15) &  -102&(10)&    \\   %
78&35 & -65&(15)  &  -58&(9)  &   \\   %
\hline \hline
\end{tabular}
\end{center}
\end{table}

Our numerical calculations use the collision model built in
Ref.~\cite{pra08-simoni}. The short-range molecular potentials are
parameterized in terms of singlet and triplet $s$-wave scattering
lengths, $a_s$ and $a_t$.  The long range interatomic interaction is
expanded in a multipole series in terms of $C_6$, $C_8$ and $C_{10}$
dispersion coefficients.  Relativistic interactions are relatively
weak for alkali species. Inclusion of the the dipolar interaction
between the atomic electron spins and of the second-order spin-orbit
interaction~\cite{Mies} is indeed sufficient to explain the available
data on the isotopic pairs.

Here, we vary the singlet and triplet $s$-wave
scattering lengths and the van der Waals coefficient $C_6$ until a
good agreement with the data is found. The additional potential
parameters are kept fixed to the values of Ref.~\cite{pra08-simoni}.
The energy levels included in the data are known to be molecular
states with null orbital angular momentum $\ell^\prime$. The
magnetic field location at which these levels become degenerate with
the energy of the separated atoms corresponds to the two $s$-wave
resonances observed in \cite{prl08-gregor}.

Note that in our fitting minimization procedure the magnetic field is
the independent variable and it assumed without error. The actual
experimental uncertainty on $B$ is accounted for by projecting it on the energy
axis and adding it in quadrature to the direct measurement. As it can
be remarked from the data, the level corresponding to the 79\,G
resonance varies rapidly with magnetic field, such that a small $B$
uncertainty of $0.05$\,G amounts to a $\sim 15$\,kHz error on the
molecular state energy. Conversely, the level associated with the
38\,G resonance varies more slowly with $B$ and the error is
determined by the direct energy measurement only.

We include in the data set two newly observed narrow features and the
narrow $s$-wave resonance of~\cite{pra08-weber}, whose position can be
very precisely determined.  Magnetic resonances of different nature
have been observed in cold atom collisions in a number of
homonuclear~\cite{papers1} and heteronuclear
systems~\cite{pra06-ferlaino, papers2}. In general, broad resonances
result from spin-exchange coupling of incoming $s$-wave atomic pair with
$\ell^\prime=0$ molecules. Such processes are induced by the
spherically symmetric exchange interaction and conserve the orbital
angular momentum $\ell$, its projection $m$ and the projection $m_f$
of the total hyperfine spin of the system in the direction of the
applied magnetic field.

Relativistic spin-spin and second order spin-orbit interactions are
anisotropic and enable first order coupling of $\ell$-wave atomic
pairs with $\ell^\prime$ symmetry molecules, provided that
$|\ell^\prime - \ell| \leq 2$, the $\ell=0 \to \ell^\prime=0$
transitions being forbidden. Rotational symmetry about the external
magnetic field implies exact conservation of the total angular
momentum projection $m+m_f$. The total interaction couples incoming
$s$-wave atomic pairs with $m_f=2$ to states characterized by
$\ell^\prime=0,2$ and $0 \leq m_f \leq 4$. Higher order couplings to
different $\ell^\prime$ and $m_f$ are possible but only give rise to much
narrower features.

Collisions for atoms incoming in $\ell>0$ partial waves tend to be
suppressed at temperatures below the $\ell$-wave centrifugal
barrier. We take into account spin-exchange $p$-wave resonances,
characterized by $\ell^\prime=1$ and $m_f=2$. Such resonances have
been observed even at very low temperature \cite{pwave}. Finally,
$p$-wave resonances induced by spin-spin coupling are in principle
possible but they can be expected to be even weaker because of the
weak coupling strength and the energy suppression. The corresponding
$m_f$ for direct coupling with incoming $p$-waves is bounded to $0
\leq m_f \leq 4$.

\begin{figure}[tbp] 
  \begin{center}
    \includegraphics[width=.8\columnwidth,clip]{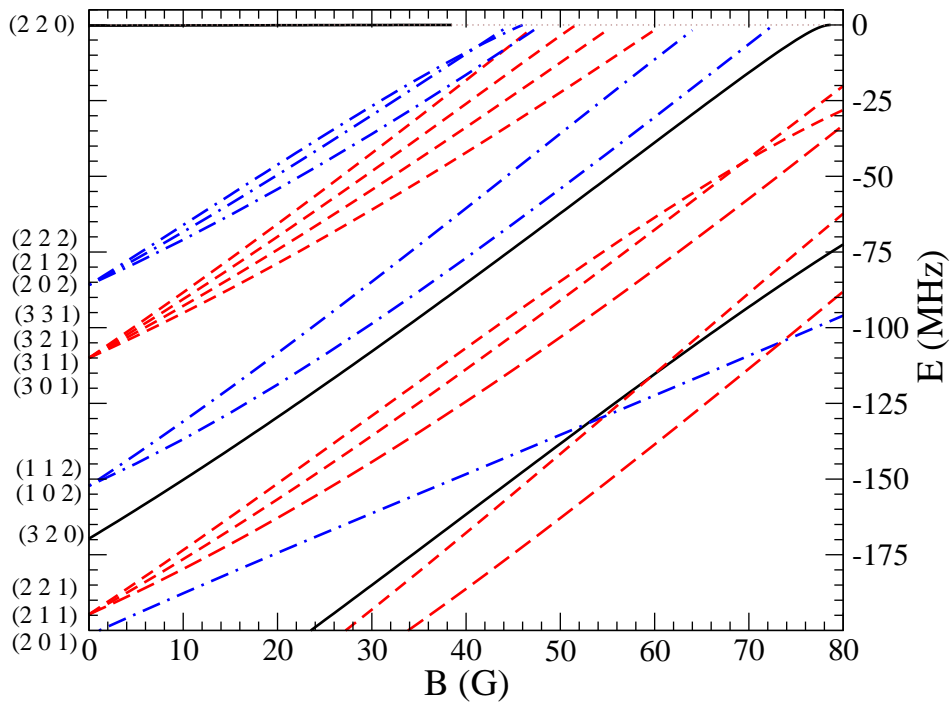}
    \caption{Molecular level scheme of \kq\rb\ with $m_f= 0 -4$ and
      $\ell^\prime=0$ (full line), 1 (dashed line), 2 (dash-dotted
      line).  The quantum numbers $(f m_f \ell^\prime)$ are also
      shown. All energies are referred to the atomic threshold, i.e.,
      the energy of a pair of unbound atoms far apart.}
\label{fig:levels}
\end{center}
\end{figure}

The model of Ref. \cite{pra08-simoni} allows us to give a tentative
assignment of the data. Towards this aim, it is instructive to
calculate the near-threshold molecular levels for the lowest values of
orbital angular momentum $\ell^\prime=0,1,2$ (see
\fref{fig:levels}). A simplified approach that neglects the spin
interactions is fully adequate for this purpose, since the resulting
level shifts are small. In this approximation $m_f$, $\ell$ and $m$
are good quantum numbers.  The $f$ quantum number is precisely defined
only when $B=0$. Thus, we use it as an approximate quantum label for
the levels in the low-field domain $B<80$\,G of \fref{fig:levels}. We
only show levels coupled in first order to incoming $s$- or $p$-wave
atoms.

\Fref{fig:levels} allows us to identify the molecular levels
associated with the two observed loss features of \fref{fig:ddwave}.
The one at lower field is univocally associated with a $\ell^\prime=2$
molecular level.  The only possible ambiguity for the higher field
feature arises from the crossing of a $\ell^\prime=1$ level with $m_f=3$ very
close to the observed value. This possibility is excluded since this
feature would be a spin-spin induced $p$-wave resonance. If such a
weak feature were observable, one should then also observe a stronger
spin-exchange $p$-wave loss feature at approximately 50\,G, which is
not detected. Moreover, with the present resolution one should observe
the typical doublet structure of $p$-wave peaks \cite{bohn}.  We
conclude that the molecular level associated with the $47$~G resonance
has also $\ell^\prime=2$ angular momentum.

With this assignment, we optimize our collisional model to obtain the
best-fit parameters: 
\bea
a_s&=&-109(2)a_0  \nonumber \\
a_t&=&-213(3) a_0 \nonumber \\
C_6&=&4285(8) E_h a_0^6.  
\eea 
where $a_0$ denotes the Bohr radius and $E_h$ the Hartree energy.

The reduced weighted $\chi^2$ is $0.92$ and the maximum deviation
between theoretical and empirical values is about two standard
deviations. The present values are in agreement with
Ref~\cite{pra08-simoni}. In fact, if the bound states and the
resonances in the three isotopic pairs are simultaneously fit, the
following model potential parameters are determined: \bea
a_s&=&-109.6(2)  a_0  \nonumber \\
a_t&=&-213.6(4) a_0 \nonumber \\
C_6&=&4288(2)  E_h a_0^6  \nonumber \\
C_8&=&4.76(5)\times 10^5  E_h a_0^8  \nonumber \\
A_{\rm ex}&=&2.01(4)\times 10^{-3} E_h.  \nonumber \\
\eea At variance with \cite{pra08-simoni} the strength $A_{\rm ex}$ of
the exchange interaction \cite{pra05-stwalley} is also included in the
fit parameters.  The quality of the fit is similar to the one of
\cite{pra08-simoni} with a reduced $\chi^2$ of $1.1$ and a maximum
discrepancy with the experimental data of at most two standard
deviations. The theoretical data do not show any systematic positive
or negative shift with respect to the observed features. This
strengthens the conclusion obtained in Ref.~\cite{pra08-simoni} that
breakdown of the Born-Oppenheimer approximation \cite{Tiemann} does
not produce measurable effects at the current level of precision.

\section{Conclusions}\label{sec:conclusions}
We have shown that recently published data on rf association of
Feshbach dimers and newly determined Feshbach resonances in higher
order partial waves lead to an improved collisional model for
\kq\rb. Such a collisional model is instrumental for future
experiments on this mixture. In particular, KRb dimers appear among
the best candidates to produce a molecular BEC with dipolar
interactions. Indeed, bosonic KRb should be amenable to efficient
transfer toward the rovibrational ground state with a single Raman
pulse, following the pioneering demonstration on their fermionic
\kqa\rb\ counterparts \cite{science08-jin}. The route to stable
bosonic molecules presumably will take advantage of dimer association
starting from atoms trapped in optical lattices. As already recognized
in \cite{prl03-zoller}, a double Mott insulator with unit filling (per
species) will enable the creation of molecules secluded in individual
lattice sites, thereby shielded from collisions with both unbound
atoms and other molecules \cite{prl06-gregor}. In addition, the
lattice is an essential tool for implementing several models of bosons
with long range interactions \cite{prl02-lewenstein}.

\ack We acknowledge L De Sarlo and G Varoquaux for their contribution
to the early stage of the experiment and all the members of the
Quantum Degenerate Gases group in Florence for discussions. Financial
support has been provided by CNR (EuroQUAM DQS, QUDIPMOL), Ente Cassa di
Risparmio di Firenze, EU (STREP CHIMONO, NAMEQUAM) and INFN
(SQUATSuper).

\section*{Appendix}
We describe here the theoretical model we used to fit the line shape
of rf association spectrum, i.e., of the total atom number remaining
after a rf pulse of fixed length as a function of the rf
frequency. Despite several simplifying approximations, the model
captures the main features of the process, such as the broadening and
the asymmetry of the spectrum. It is worth mentioning that the very
same model was summarily described and applied in the analysis of
Ref. \cite{pra08-weber}. Here we report a more detailed derivation of the
line shapes.

We consider an atom pair in the center-of-mass reference frame. We
introduce the annihilation operators for atoms A with momentum $p$,
atoms B with momentum $-p$ and dimers with zero momentum:
$\hata,\hatb$ and $\hatm$. In the absence of rf coupling, the
time-dependent free hamiltonian is $\hat{H}_0=(-\ebind
(B(t))-i\gamma/2) \hatm^\dagger\hatm + p^2/(2 m_A)\hata^\dagger\hata+
p^2/(2 m_B) \hatb^\dagger\hatb$, where $m_A$ and $m_B$ denote the
atomic masses and the time-dependence of the binding energy is brought
by the dependence of the (positive-defined) binding energy $\ebind$ on
the oscillating Feshbach field $B(t)=B_{\rm dc}+b\sin (\omega t)$. The
molecules have a finite lifetime $\gamma^{-1}$, due to relaxation
processes.

The coupling is due to the oscillating part of the Feshbach
field. Following Ref.~\cite{pra07-molmer}, we consider a
single-particle coupling strength proportional to the derivative of
the oscillating field $\Omega(B(t))=\Omega_0(B_{\rm dc})\cos(\omega
t),$ with $\Omega_0(B):=\omega b \left( \frac{\partial}{\partial B'}
  \langle \psi_m (B)|\psi(B')\rangle \right)_{B'=B}$. Thus the rf
coupling hamiltonian is given by $\hat{H}_{\rm rf}= \Omega (t)
(\hatm^\dagger\hata\hatb + \hata^\dagger\hatb^\dagger\hatm)$. With
this, we write the Heisenberg equation for the operators
$\hata,\hatb,\hatm$ and take the expectation values while neglecting
all correlations. In practice, we replace the above operators with
$c$-numbers $\alpha,\beta,\mu$:
\begin{eqnarray}
  i\dot \mu &=& \left(-\ebind(t)/\hbar - i\gamma/2\right)\mu
  + \Omega_0\cos(\omega t) \alpha\beta  
  \nonumber \\
  i\dot \alpha &=& p^2/(2\hbar m_A)\,\alpha+\Omega_0\cos(\omega t) 
 \beta^\ast \mu   
  \label{eq:nlbloch}\\
  i\dot \beta &=& p^2/(2\hbar m_B)\,\beta+\Omega_0\cos(\omega t) 
 \alpha^\ast \mu .  
  \nonumber
\end{eqnarray}

As expected from the conservation of total number of particles, the
quantity $2|\mu|^2+|\alpha|^2+|\beta|^2$ is constant for $\gamma=0$.

We can solve analytically this set of non-linear Bloch equations,
provided we introduce {\it (i)} the rotating-wave approximation,
i.e., we write the equations in terms of the slowly varying amplitudes
$\tilde\alpha:=\alpha \exp(i p^2/(2\hbar m_A)t),\, \tilde\beta:=\beta
\exp(i p^2/(2\hbar m_B) t),\,\tilde\mu:=\mu\exp(-i(\omega
-\ekin/\hbar)t)$, with $\ekin=p^2/(2m_A)+p^2/(2m_B)$, and neglect all
rapidly oscillating terms; {\it (ii)} the adiabatic approximation,
whereby we consider that, since the molecular decay rate is fast, we
can take the molecular amplitude $\tilde\mu$ to adiabatically follow
the evolution dictated by the slower rf coupling, in practice ${\rm
  d}\tilde\mu/{\rm d}t =0$. As a consequence, we have:
\begin{eqnarray}
  \tilde \mu &=& -\frac{\Omega_0}{2(\delta-i\gamma/2)}\tilde\alpha\tilde\beta
  \nonumber \\
  \dot N_A&:=& \dot\alpha \alpha^\ast + {\rm c.c.} = -\frac{\gamma\Omega_0^2}{4\delta^2+\gamma^2}
  N_A N_B
  \label{eq:nlbloch2}\\
  \dot N_B&:=& \dot\beta \beta^\ast + {\rm c.c.} = -\frac{\gamma\Omega_0^2}{4\delta^2+\gamma^2} 
  N_A N_B
  \nonumber
\end{eqnarray}
with $\delta:=\omega-\ebind/\hbar-\ekin/\hbar$.

Obviously the difference $n:=N_A-N_B$ stays constant, while the sum
$\ntot:=N_A+N_B$ obeys the equation
\begin{equation}
  \dot \ntot = -\gammaeff (\ntot^2-n^2), 
  \quad \gammaeff:=\gamma\frac{\Omega_0^2}{2(4\delta^2+\gamma^2)}
\end{equation}
that has the following solution
\begin{equation}
  \label{eq:ntot}
  \ntot(t)=n \frac{C\exp(2 n \gammaeff t)+1}{C\exp(2 n \gammaeff t)-1}, 
  \quad C:=\frac{\ntot (0) + n}{\ntot(0)-n}
\end{equation}
For $n=0$, the above reduces to 
\begin{equation}
  \label{eq:simplentot}
 \ntot(t)=\frac{\ntot(0)}{1+\ntot(0) \gammaeff t} 
\end{equation}
which is the well-known solution of the rate equation $\dot
N=-\gammaeff N^2$, that is used to describe losses due to two-body
collisions.

The solution \eref{eq:ntot} depends on the atomic kinetic energy
$\ekin$, whose values are spread over a range of the order of the
temperature $T$. Under our experimental conditions, the thermal
distribution of kinetic energies dominates the broadening of the rf
spectra over the linewidth dictated by the decay rate $\gamma$. In
order to compare with the measured spectra, we need to take a thermal
average of \Eref{eq:ntot} by convolving with the Boltzmann
distribution of kinetic energies proportional to
$\sqrt{\ekin}\exp(-\ekin/k_B T)$. The resulting line shape describes
well the strong asymmetry of the observed spectra.

A more detailed inspection of Equations \eref{eq:nlbloch} reveals
another interesting feature: Molecular association also occurs for
fractional frequencies of the resonant frequency (see
\fref{fig:rfline}). This is experimentally observed and confirmed  by
the results of numerical integration of Equations \eref{eq:nlbloch}.

Indeed the fact that the binding energy is modulated at the rf
frequency $\omega$ makes it possible that the molecular amplitude
contains Fourier components oscillating at integer multiples of
$\omega$. When one of these harmonics is close to the time-averaged
value of the binding energy, the transfer of population to the
molecular level occurs.

Another important experimental finding confirmed by the numerical
simulations is the shift of the resonant peak frequency occurring when
the binding energy depends nonlinearly on the magnetic field detuning
from the Feshbach resonance. This is also easily understood when we
consider that the time-averaged binding energy is shifted from the
value in the absence of modulation. If the binding energy is quadratic
with the magnetic field detuning $\ebind=\eta \Delta B^2$, which is
the case next to a Feshbach resonance, we have that
$\langle\ebind\rangle_t=\ebind (\Delta B_{\rm dc})+\eta b^2/2$, since
$\Delta B(t)=\Delta B_{\rm dc}+b\sin(\omega t)$. This relationship has
been experimentally verified \cite{pra08-weber}.


\section*{References}
\begin{thebibliography}{10}

\bibitem{prl02-pillet} Vanhaecke N, de Souza Melo W, Tolra B L,
  Comparat D and Pillet P 2002 Phys. Rev. Lett. {\bf 89} 063001

\bibitem{science02-heinzen} 
  Wynar R, Freeland R S, Han D J, Ryu C and Heinzen D J 2002 Science
  {\bf 287} 1016

\bibitem{science03-grimm} Jochim S, Bartenstein M, Altmeyer A, Hendl G,
  Riedl S, Chin C, Hecker Denschlag J, Grimm R 2003 Science {\bf 302}
  2101

\bibitem{nature03-greiner} Greiner M, Regal C A and Jin D S 2003
  Nature {\bf 426} 537

\bibitem{prl06-sengstock} Ospelkaus C, Ospelkaus S, Humbert L, Ernst
  P, Sengstock K and Bongs K 2006 \prl\ {\bf 97} 120402

\bibitem{prl04-gndmot} Mancini M W, Telles G D, Caires A R L,Bagnato V
  S and Marcassa L G 2004 \prl\ {\bf 92} 133203

\bibitem{science08-jin} Ni K -K, Ospelkaus S, de Miranda M H G, Pe'er
  A, Neyenhuis B, Zirbel J J, Kotochigova S, Julienne P S, Jin D S and Ye
  J 2008 Science {\bf 322} 231

\bibitem{pra08-weber} Weber C, Barontini G, Catani J, Thalhammer G,
  Inguscio M and Minardi F 2008 \pra\ {\bf 78} 061601(R)

\bibitem{prl09-dieckmann} Voigt A -C, Taglieber M, Costa L, Aoki T, Wieser W,
  H\"ansch T W and Dieckmann K 2009 \prl\ {\bf 102} 020405

\bibitem{pra08-simoni} Simoni A, Zaccanti M, D'Errico C, Fattori M,
  Roati G, Inguscio M and Modugno G 2008 \pra\ {\bf 77} 052705

\bibitem{pra06-ferlaino} Ferlaino F, D'Errico C, Roati G, Zaccanti M,
  Inguscio M, Modugno G and Simoni A 2006 \pra\ {\bf 73} 040702(R)

\bibitem{pra07-arlt} Klempt C, Henninger T, Topic O, Will J, Ertmer W,
  Tiemann E and Arlt J 2007 \pra\ {\bf 76} 020701(R)

\bibitem{pra06-catani} Catani J, De Sarlo L, Maioli P, Minardi F and
  Inguscio M 2006 \pra\ {\bf 73} 033415

\bibitem{pra07-desarlo} De Sarlo L, Maioli P, Barontini G, Catani J,
  Minardi F and Inguscio M 2007 \pra\ {\bf 75} 022715

\bibitem{prl08-gregor} Thalhammer G, Barontini G, De Sarlo L, Catani
  J, Minardi F and Inguscio M 2008 \prl\ {\bf 100} 210402

\bibitem{prl05-wieman} Thompson S T, Hodby
  E and Wieman C E 2005 \prl\ {\bf 95} 190404

\bibitem{Mies} Mies F H, Williams C J and Julienne P S,
J. Res. Natl. Inst. Stand. Technol. 1996 {\bf 101} 521

\bibitem{papers1} Leo P J, Williams C J and Julienne P S 2000 \prl\
  {\bf 85} 2721; van Kempen E G M, Kokkelmans S J J M F, Heinzen D J
  and Verhaar B J 2002 \prl\ {\bf 88} 093201; D'Errico C, Zaccanti M,
  Fattori M, Roati G, Inguscio M, Modugno G and Simoni A 2007 New
  J. Phys {\bf 9} 223

\bibitem{papers2} Burke J P, Bohn J L, Esry B D and Greene C H 1998
  {\bf 80} 2097; Gacesa M, Pellegrini P and C{\^o}t{\'e} R 2008 \pra\
  {\bf 78}, 010701(R); Wille E, Spiegelhalder F M, Kerner G, Naik D,
  Trenkwalder A, Hendl G, Schreck F, Grimm R, Tiecke T G, Walraven J T
  M, Kokkelmans S J J M F, Tiesinga E and Julienne P S 2008 \prl\
  {\bf 100} 053201

\bibitem{pwave} Schunck C H, Zwierlein M W, Stan C A, Raupach S M F,
  Ketterle W, Simoni A, Tiesinga E, Williams C J and Julienne P S 2005
  \pra\ {\bf 71} 045601

\bibitem{bohn} Ticknor C, Regal A C, Jin D S and Bohn J L 2004
  \pra\ {\bf 69} 042712

\bibitem{pra05-stwalley} Zemke W T, C{\^o}t{\'e} R and Stwalley W C
  2005 \pra\ {\bf 71} 062706

\bibitem{Tiemann} Falke S, Tiemann E and Lisdat C 2007 \pra\ {\bf 76} 012724

\bibitem{prl03-zoller} Damski B, Santos L, Tiemann E, Lewenstein M,
  Kotochigova S, Julienne P and Zoller P 2003 \prl\ {\bf 90} 110401

\bibitem{prl06-gregor} Thalhammer G, Winkler K, Lang F, Schmid S,
  Grimm R and Hecker Denschlag J 2006 \prl\ {\bf 96} 050402

\bibitem{prl02-lewenstein} G\'oral K, Santos L and Lewenstein M 2002
  \prl\ {\bf 88} 170406

\bibitem{pra07-molmer} Bertelsen J F and M\"olmer K 2007 \pra\
  {\bf 76} 043615

\end{thebibliography}
\end{document}